\documentclass[12pt,a4paper,twoside,fleqn]{article}
\usepackage{calc}
\usepackage{graphicx}
\usepackage{caption2}
\usepackage{amssymb}   
\usepackage{subfigure}

\newcommand{\barre}[1]{%
	\setbox1=\hbox{$#1$} \dimen2=\ht1 \dimen3=\dp1 \dimen4=\wd1
	\setbox2=\hbox{\sl /}
	\dimen1=\wd1 \advance\dimen1 by -\wd2 \divide\dimen1 by 2
	\advance\dimen1 by \wd2 \advance\dimen1 by 0.4pt
	\setbox3=\hbox to \wd1{\hss \box1 \kern -\dimen1 \box2\hss}
	\ht3=\dimen2 \dp3=\dimen3 \wd3=\dimen4
	\box3
	}
 
\newcommand{\beq}{\begin{equation}}
\newcommand{\eeq}{\end{equation}} 
\newcommand{\bea}{\begin{eqnarray}}
\newcommand{\eea}{\end{eqnarray}} 
\newcommand{\nn}{\nonumber}
\newcommand{\ba}{\begin{array}} 
\newcommand{\ea}{\end{array}}

\newlength{\centermargin}
\newlength{\bordermargin}
\newlength{\docwidth}
\newlength{\docheight}

\newlength{\oldtopmargin}
\newlength{\oldabovecaptionskip}
\newlength{\epswidth}
\newenvironment{xmgrfigure}
{\setlength{\abovecaptionskip}{-2pt} \begin{figure} }
{\end{figure}   \setlength{\abovecaptionskip}{\oldabovecaptionskip} }

\begin{document}

\begin{titlepage}

\title{Quark and gluon fragmentation functions into photons.}
\author{L. Bourhis$^{a}$ \and M. Fontannaz$^a$ \and J. Ph. Guillet$^b$}
\date{}

\maketitle

\begin{centering}
$^a$ Laboratoire de Physique Th\'eorique et Hautes
Energies\footnote{Laboratoire associ\'e au CNRS - URA 0063.}\\
Universit\'e de Paris XI, b\^atiment 211\\ 
F-91405 Orsay Cedex, France\\
$^b$ Laboratoire de Physique Th\'eorique ENSLAPP\footnote{URA 14-36 du
CNRS, associ\'ee \`a l'Ecole Normale Sup\'erieure de Lyon, et au
Laboratoire d'Annecy-le-Vieux de Physique des Particules.} - Groupe
d'Annecy\\ 
LAPP, IN2P3-CNRS, B.P. 110\\
F-74941 Annecy-le-Vieux Cedex, France\\
\end{centering}

\vspace{5em}

\begin{abstract}

The fragmentation functions of quarks and gluons into photons are studied beyond the Leading Logarithm approximation. We address the nature of the initial conditions of the evolution equation solutions and study problems related to factorization scheme invariance. The possibility of measuring these distributions in LEP experiments is discussed, and a comparison with existing data is made.

\end{abstract}
\vspace{2em}
LPTHE-ORSAY 96/103\\
ENSLAPP-A-631/96
\end{titlepage}

\section{Introduction}

The fragmentation of quarks and gluons into photons can be observed in $e^+e^-$ annihilation experiments and in the production of large-$p_{\bot}$ photons in hadronic collisions. This phenomenom is described by the distributions $D_q^{\gamma}(z, M^2)$ and $D_g^{\gamma}(z, M^2)$ where $z$ is the fractional momentum carried away by the photon and $M^2$ a time-like scale fixed by the hard process ($M^2 \sim Q^2$ in $e^+e^-$-annihilation and $M^2 \sim p_{\bot}^2$ for large-$p_{\bot}$ photons). Unlike the fragmentation into hadrons which are complex bound-states, the photon has a known pointlike coupling to the quark. Therefore we expect these distributions to be fully calculable in perturbative QCD. Witten was the first to show that this is indeed the case [1], at least for $M^2$ large enough to neglect non-perturbative effects, and he wrote the Leading Logarithm (LL) expressions for $D_a^{\gamma}(z, M^2)$ $(a = q, g)$. In practice, it turns out that we need to know the fragmentation functions in kinematical domains where $M^2$ is not asymptotically large ($M^2 \sim p_{\bot}^2 \sim 25\ GeV^2$ in fixed-target direct-photon experiments). These non-perturbative contributions and Beyond Leading Logarithm corrections (BLL) to Witten's LL results are sizeable. It is the purpose of this paper to present a careful analysis of these effects and a new parametrization of $D_q^{\gamma}(z, M^2)$ and $D_g^{\gamma}(z, M^2)$.
  
New experimental results justify the updating of an analysis published some years ago [2]. Since then the LEP collaborations studied in detail the quark fragmentation into isolated hard photons; the inclusive fragmentation functions $D_a^{\gamma}(z, M^2)$ should also be measurable [3]. On the other hand new precise data on direct photon production at large $p_{\bot}$ has been [4], or will [5] be presented soon, requiring more precise theoretical inputs. Finally it is now possible to better constrain the non-perturbative part of the fragmentation functions which is obtained by means of the Vector Meson Dominance Model (VDM). Indeed, new data on the inclusive $\rho$ production  at LEP  \cite{delphi,aleph} allow a better control of this contribution. 

The study of the fragmentation functions follows the theoretical approach developed in the analysis of the crossed reaction, namely the Deep Inelastic Scattering of a virtual photon on a real photon, which probes the parton contents of a real photon [8]. The good agreement between theory and data obtained in this channel lets us hope to derive sound predictions for the $D_a^{\gamma}(z, M^2)$ distribution. Here we study only the inclusive fragmentation functions, without any isolation condition around the photon. The isolated case [9] raises theoretical problems concerning the Infra-Red stability of the prediction [10, 11] that we do not discuss in this paper.

\section{Theoretical Background}

The fragmentation functions $D_a^{\gamma}(z, M^2)$ verify the inhomogeneous DGLAP (Dokshitzer, Gribov, Lipatov, Altarisi, Parisi) equations [12, 13] (the convolution $f \otimes g(z)$ is defined by $\int_0^1 du \  dv \ f(u) \ g(v) \delta (uv - z)$)
\beq
M^2 {\partial D_{ns, i}^{\gamma} \over \partial M^2} = C_{ns,i} \
K_{\gamma q} + P_{ns} \otimes D_{ns,i}^{\gamma} 
\label{DGLAP NS}
\eeq
\noindent for the non-singlet sector ($C_{ns,i} = 2(e_i^2 - <e_i^2>)$), and
\bea
M^2 {\partial D_q^{\gamma} \over \partial M^2} = C_s \ K_{\gamma q} + P_{qq} 
\otimes D_q^{\gamma} + P_{gq} \otimes D_g^{\gamma} \nonumber \\
M^2 {\partial D_g^{\gamma} \over \partial M^2} = C_s \ K_{\gamma g} + P_{qg}
\otimes D_q^{\gamma} + P_{gg} \otimes D_g^{\gamma} 
\label{DGLAP sing}
\eea
\noindent for the singlet sector ($C_s = 2 N_f <e_i^2>$), with $D_q^{\gamma} = \sum\limits_{i=1}^{N_f} (D_{q_i}^{\gamma} + D_{\bar{q}_i}^{\gamma})$ and $D_{ns,i}^{\gamma} = (D_{q_i}^{\gamma} + D_{\bar{q}_i}^{\gamma}) - D_q^{\gamma}/N_f$. The inhomogeneous kernels have a perturbative expansion
\beq
K_{\gamma a}(z, M^2) = {\alpha \over 2 \pi} \left ( K_{\gamma
a}^{(0)}(z) + {\alpha_s
\over 2 \pi} (M^2) \ K_{\gamma a}^{(1)}(z) + \cdots \right ) 
\label{Kgamma}
\eeq
\noindent as do the homogeneous kernels $P_{ab}$. The kernels $K_{\gamma a}$ are given in [2], and the homogeneous ones can be obtained from  [14]. Let us notice that the coupling of the gluon to the photon can only take place through a quark loop~; therefore the expansion (\ref{Kgamma}) of $K_{\gamma g}$ starts at order $O(\alpha_s)$. 

In the moment space ($f(n) = \int_0^1 dz \ z^{n-1} f(z)$), (\ref{DGLAP NS}) and (\ref{DGLAP sing}) can easily be solved [15,16,17]. For instance for the non-singlet distribution we obtain
\beq
D^{\gamma , AN} (n , M^2) = C \int_{\alpha_s(M_0^2)}^{\alpha_s(M^2)} {d \lambda
\over \beta ( \lambda )} K_{\gamma q}(n) \ 
\hbox{\large e}\,^{\int_{\lambda}^{\alpha_s(M^2)} \frac{d \lambda '}
{\beta (\lambda ')} P(n)} 
\label{anomalous sol}
\eeq 
\noindent where we have dropped the suffixes $ns$ and $i$. The suffix $AN$ means anomalous, a qualifier given by Witten to the solutions of eq. (\ref{DGLAP NS}),(\ref{DGLAP sing}) in order to characterize their asymptotic behaviours. Indeed, with the definitions $r = \alpha_s(M^2)/\alpha_s(M_0^2)$ and $d_n = 2P^{(0)}(n)/\beta_0$, the solution (\ref{anomalous sol}) can be written, in the LL approximation,
\beq
D^{\gamma ,AN}(n, M^2) = {4 \pi \over \alpha_s(M^2)} \ {\alpha \over 2 \pi} \ 
{C \over \beta_0} \ {K_{\gamma q}^{(0)}(n) \over 1 - d_n} \left ( 1 - r^{1- d_n}
\right ) 
\label{anomalous sol LL}
\eeq
\noindent an expression which explicitly displays an asymptotic behavior proportional to $\ln\frac{M^2}{\Lambda^2}$ (in (\ref{anomalous sol LL}) we kept only the lowest order term of the $\beta$-function~: 
\[M^2\partial\alpha_s/\partial M^2 = \beta (\alpha_s) = - \alpha_s (\beta_0
\alpha_s/4 \pi + \beta_1 (\alpha_s /4 \pi )^2 + \cdots )\] 

Expression (\ref{anomalous sol}) is not the full solution of the inhomogeneous equation (\ref{DGLAP NS})~; we can add to (\ref{anomalous sol}) a general solution of the homogeneous equation (eq. (\ref{DGLAP NS}) with $K_{q\gamma} = 0$), so that the full solution is
\beq
D^{\gamma}(n, M^2) = D^{\gamma , AN}(n, M^2) + D^{\gamma ,NP}(n, M^2). 
\label{full sol}
\eeq

The physical interpretation of expressions (\ref{anomalous sol}) and (\ref{full sol}) is the following~: $D^{\gamma}(n, M^2)$ is given by the sum of a perturbative component $D^{\gamma ,AN}$ and of a non-perturbative component $D^{\gamma ,NP}$. $D^{\gamma ,AN}$ is fully calculable in perturbative QCD, as long as $M^2$ is large enough, $M^2 > M_0^2$ where $M_0^2$ is the boundary between the perturbative and non-perturbative domain. For $M^2 = M_0^2$, the perturbative approach is no longer valid and $D^{\gamma}$ is given by a non-perturbative fragmentation function $D^{\gamma ,NP}$, which verifies for $M^2 > M_0^2$ the homogeneous DGLAP equations.

The non-perturbative input $D^{\gamma , NP}(n, M_0^2)$ is not known. We modelize it following VDM and we assume (for $M^2 \leq M_0^2$) that quarks and gluons first fragment into vector mesons which then turn into photons. Therefore we could write 
\beq D^{\gamma ,NP}(n,
M_0^2) = \alpha \sum_{v=\rho, \omega , \phi} C_v \ D^v(n, M_0^2) 
\label{NP input}
\eeq
\noindent where the fragmentation functions $D^v$ may be measured in $e^+e^-$-annihilation experiments. The coefficients $C_v$ are fixed by VDM. The value of $M_0^2$ is not known, but should be of the order of an hadronic mass and we take $M_0^2 \simeq m_{\rho}^2 \simeq 0.5\ GeV^2$. The same approach in the crossed channel $\gamma \gamma^* \to X$ leads to predictions in good agreement with data [8]. 

However the approach just described is too naive as it is based on a LL approximation. At BLL order, the decomposition (\ref{full sol}) is not factorization scheme invariant, and our VDM assumption (\ref{NP input}) must be refined. Let us study this problem which is related to BLL corrections to the LL expression (\ref{anomalous sol LL}). 

\section{Non-Perturbative Input and Factorization \\Scheme}

We consider the one-photon inclusive cross-section in $e^+e^-$-annihilation. It is given by the convolution between the hard sub-process cross-sections $C_a(z)$ and the parton fragmentation functions [18]
\bea 
{1 \over \sigma_0} \ {d\sigma^{\gamma} \over dz} &=& \sum_{i=1}^{N_f} \ e_i^2  
\ C_q \otimes \left ( D_{q_i}^{\gamma}(Q^2) +
D_{\bar{q}_i}^{\gamma}(Q^2)  \right ) + 2 \sum_{i=1}^{N_f} e_i^2 \ C_g \otimes
D_g^{\gamma}(Q^2)  \nn \\
&&\mbox{}+ 2 \sum_{i=1}^{N_f}\  e_i^4 \ C^{\gamma}
\label{cross section} 
\eea 
\noindent where $\sigma_0 = 4 \pi \alpha^2/Q^2$. The hard cross-sections $C_q(z)$ and $C_g(z)$, which have expansions in powers of $\alpha_s(Q^2)$ 
\beq
C_a = \delta_{a,q} + {\alpha_s \over 2 \pi} (Q^2) \ C_a^{(1)}(z) +
\cdots 
\eeq
\noindent also appear in the one-hadron inclusive cross-sections and have been calculated in  [19] (in (\ref{cross section}) we consider the sum of the transverse and longitudinal cross-sections). $C^{\gamma}(z)$ is characteristic of reactions involving photons and describes the direct coupling of the photon to quarks in $e^+e^-$-annihilation. Its expression in the $\overline{MS}$ Factorization Scheme (FS) can be obtained from  [19]~:
\beq
C^{\gamma} = {\alpha \over 2 \pi} \ {1 + (1 - z)^2 \over z} \left ( \ln(1 - z) +
2 \ln z \right ). 
\label{direct term}
\eeq

Actually it is well known that the fragmentation functions and hard cross-sections are not univocally defined. For instance, a part of $C^{\gamma}$ can be absorbed in the fragmentation functions, leading to a new photonic FS to which correspond new functions $\widetilde{D}_a$ and $\widetilde{C}^{\gamma}$. Each term on the rhs of (\ref{cross section}) is therefore FS dependent, but the sum is not, being a physical quantity. It is easy to verify that this implies that $D^{\gamma,NP}$ in expression (\ref{full sol}) is not FS invariant. Such ambiguities appear also in the definition of coefficients $C_q$ and $C_g$ and kernels $P_{ab}$. The influence of this hadronic FS on the fragmentation functions was studied in ref [8]. In this article, we will focus only on the difficulties related to the photonic FS.  

In order to grasp this point more clearly, we calculate (\ref{anomalous sol}) including BLL corrections. Expanding in powers of $\alpha_s$, we obtain for expression (\ref{full sol})
\bea
D^{\gamma}(n, M^2) &=& {\alpha \over 2 \pi} \ {C \over \beta_0} 
K_{\gamma q}^{(0)}(n) \left \{  {4\pi \over \alpha_s(M^2)} 
\ {1 - r^{1-d_n} \over 1 - d_n} \right. \nonumber \\
&&\mbox{}+ 2 \left ( {\beta_1 \over 2 \beta_0} d_n - {2 P^{(1)}(n) \over \beta_0}
\right )  {1 - r^{1-d_n} \over 1 - d_n}   \nonumber \\
&& \left.\mbox{} + 2 \left ( (1 + d_n)
{\beta_1 \over 2 \beta_0} - {K^{(1)}_{\gamma q}(n) \over K_{\gamma q}^{(0)}(n)} 
- 2 {P^{(1)}(n) \over \beta_0} \right ) {1 - r^{-d_n} \over d_n}  
\right \} \nonumber \\
&&\mbox{}+ r^{-d_n} \ D^{\gamma ,NP}(n, M_0^2). 
\eea 
\noindent By combining this result with (\ref{cross section}) and keeping the relevant terms proportional to $(1 - r^{-d_n})$ and $r^{-d_n}$, we easily obtain (for the non-singlet contribution and writing again $C_{ns,i}$ instead of $C$) 
\bea
{1 \over \sigma_0}{d\sigma^\gamma \over dz}(n) &=& \sum_i e_i^2
\left \{  
C_{ns,i}\ \frac{\alpha}{2\pi}\ \frac{2}{\beta_0}\left (
(1 + d_n)\ \frac{\beta_1}{\beta_0}\ K_{\gamma q}^{(0)}(n) - 
K_{\gamma q}^{(1)}(n)
\right. \right. \nn \\
&&\left .\mbox{}- 2\ \frac{K_{\gamma q}^{(0)}(n)}{\beta_0}\ P^{(1)}(n) \right ) 
\frac{1 - r^{-d_n}}{d_n}  
\nn \\
&& \left. \mbox{}+ C_{ns,i}\ C^{\gamma}(n) + D_{ns,i}^{\gamma,NP}(n, M_0^2)\ r^{-d_n} 
\right \} +\cdots 
\nn \\
&=& \sum_i e_i^2 \left \{ \left [ C_{ns,i} \ C^{\gamma}(n)\ d_n - C_{ns,i} 
{\alpha \over 2 \pi} {2 \over \beta_0} K_{\gamma q}^{(1)}(n) \right ] 
{1 - r^{-d_n} \over d_n} \right . \nn \\
&&\left . \mbox{}+ \left [ C_{ns,i} \ C^{\gamma}(n) + D^{\gamma ,NP}_{ns,i}(n, M_0^2) 
\right ] r^{-d_n} + \cdots \right \} + \cdots 
\label{cross section moment}
\eea

Expression (\ref{cross section moment}) explicitly shows [20] that the combinations $C^{\gamma}(n) d_n - {\alpha \over 2 \pi} {2 \over \beta_0} K_{\gamma q}^{(1)}(n)$ and $C_{ns,i} \ C^{\gamma}(n) + D^{\gamma ,NP}_{ns,i}(n)$ are FS invariant~; if we change the scheme, we must obtain
\beq
C_{ns,i} \ \widetilde{C}^{\gamma}(n) + \widetilde{D}^{\gamma ,NP}_{ns,i}(n) =
C_{ns,i} \ C^{\gamma}(n) + D^{\gamma ,NP}_{ns,i}(n) 
\label{FS invariant}
\eeq
\noindent and we clearly see that the ``non-perturbative'' component cannot correspond to a VDM contribution alone, which should be FS invariant. 

In ref [8], we discussed the structure of $D^{\gamma ,NP}$ (actually in the DIS channel) in detail and showed that it consists of two parts. One part includes all the non perturbative effects and is scheme independent. The other part depends on the scheme and can be perturbatively calculated. It corresponds to the collinear part of $C_{\gamma}$. In this paper we quote the result without proof, refering the interested reader to the original paper [8].
\beq
D^{\gamma ,NP}_{ns,i}(z, M_0^2) = D^{\gamma ,VDM}_{ns,i}(z, M_0^2) - C_{ns,i} \
D^{\gamma ,\overline{MS}}(z)
\label{modified NP input} 
\eeq
\noindent where 
\beq
D^{\gamma ,\overline{MS}}(z) = {\alpha \over 2 \pi} \left ( {1 + (1 -
z)^2 \over z} \left( \ln (1 - z) + \ln (z) \right ) + z \right ). 
\label{MS bar input} 
\eeq

Expression (\ref{modified NP input}) is our initial condition at $M^2 = M_0^2$. A similar result can be obtained for the singlet sector with $C_{ns,i}$ replaced by $C_s$. For the gluon fragmentation, we have $D^{\gamma ,NP}_g(z, M_0^2) = D^{\gamma ,VDM}_g(z, M_0^2)$.

The previous discussion is valid for light quarks. For massive quarks, we neglect the VDM component. For instance, we neglect the $\psi$-dominance contribution to the fragmentation of a charm quark into photons. But we still have a ``non-perturbative'' input. Indeed Nason and Webber [21] calculated the fragmentation of a heavy quark or anti-quark into  a photon (actually a gluon in these calculations) with the result (dropping powers of $m_Q/Q$)
\beq
{1 \over \sigma_0} \ {d\sigma^{\gamma}_Q \over dz}(z) = e_Q^4  \
{\alpha \over 2 \pi} {1 + (1 - z)^2 \over z} \ \ln {M^2 \over m_Q^2} +
e_Q^4 \ C_Q^{\gamma} (z) 
\eeq
\noindent where the direct term $C_Q^{\gamma}$, calculated in the massive FS, is given by~:
\beq
C_Q^{\gamma}(z) = C^{\gamma}(z) - {\alpha \over 2 \pi} \ {1 + (1 - z)^2 \over z} (2 \ln z + 1). 
\eeq
\noindent $C^{\gamma}(z)$ is the direct term in the $\overline{MS}$ scheme given by (\ref{direct term}). By taking into account BLL corrections one obtains an expression similar to (\ref{cross section moment}), but in which the kernels are calculated in the massive scheme and $r$ is replaced by  $r_Q = \alpha_s(M^2)/\alpha_s(m_Q^2)$. In particular, in the terms
\beq
\frac{1}{2e_Q^4}{1 \over \sigma_0} \ {d\sigma^{\gamma} \over dz}(n) = - {\alpha \over 2 \pi} \ {2 \over \beta_0} 
K^{(1)}_{\gamma Q}(n) {1 - r_Q^{-d_n} \over d_n} + C_Q^{\gamma}(n) + \cdots 
\label{heavy cross section moment}
\eeq
\noindent one recognizes the FS invariant combination 
\beq
- {\alpha \over 2 \pi} {2
\over \beta_0} {K_{\gamma Q}^{(1)}(n) \over d_n} + C_Q^{\gamma}(n).
\label{heavy FS invariant}
\eeq
\noindent Expression (\ref{heavy cross section moment}) can be transformed into the $\overline{MS}$ scheme we used in this paper~: 
\bea
\lefteqn{- {\alpha \over 2 \pi} \ {2 \over \beta_0} K_{\gamma
Q}^{(1)}(n) {1 - r_Q^{-d_n} \over d_n} + C_Q^{\gamma}(n) =}
\nonumber \\
&&- {\alpha \over 2 \pi} \
{2 \over \beta_0}  \left ( K_{\gamma Q}^{(1)}(n) + \delta K(n) \right ) 
\left ( {1 - r_Q^{-d_n} \over d_n} \right ) \nonumber \\ 
&&+ \left ( C_Q^{\gamma}(n) + {\alpha \over 2 \pi} \
{2 \over \beta_0} \frac{\delta K(n)}{d_n} \right ) - {\alpha \over 2\pi} \ 
{2 \over \beta_0} \ {\delta K(n) \over d_n} r_Q^{- d_n} 
\label{heavy cross section moment 2}
\eea
\noindent with $\overline{MS}$ expressions
\beq
K_{\gamma q}^{(1)} = K_{\gamma Q}^{(1)} + \delta K
\label{heavy change 1} 
\eeq
\noindent and
\beq
C^{\gamma} = C_Q^{\gamma} + {\alpha \over 2 \pi} \ {2 \over \beta_0} \frac{\delta K}{d_n} 
\label{heavy change 2} 
\eeq
\noindent From (\ref{heavy change 1}) and (\ref{heavy change 2}) we see that we can recover the massive result (\ref{heavy cross section moment 2}) by working in the $\overline{MS}$ scheme, but with a non-zero input at $M^2 = m_Q^2$ given by
\beq
D_Q^{\gamma, \overline{MS}}(z, m_Q^2) = - {\alpha \over 2 \pi} e_Q^2 {1 + (1 - z)^2 \over z} (2 \ln z + 1). 
\label{massive input}
\eeq

Let us end this section by comparing our present approach with previous BLL studies.

The authors of ref [17] use a different approach but obtain similar results for the ``non perturbative'' input. Invoking the ``perturbative stability'', they choose to work with a factorization scheme (called $DIS_\gamma$) in which the direct term $C^\gamma_{DIS_\gamma}(z)$ vanishes (more precisely the transverse direct term). Then they assume that the input at $M^2=M_0^2$ is given simply by the VDM contribution.

Using (\ref{FS invariant}), we can translate this input in the $\overline{MS}$ language. In the non-singlet case, we find
\beq
D_{ns,i}^{\gamma,NP}=\tilde{D}_{ns,i}^{\gamma,NP}-C_{ns,i}\ C^\gamma_T=
D_{ns,i}^{\gamma,VDM}-C_{ns,i}\ C^\gamma_T
\label{disg}
\eeq
\noindent where $\sim$ now means $DIS_\gamma$ and with 
\beq
C^\gamma_T=\frac{\alpha}{2\pi} \frac{1+(1-z)^2}{z} \left ( \ln(1-z) +
2\ln z \right ) -2\frac{1-z}{z}
\label{Ctrans}
\eeq
\noindent valid for any flavour. We see that this expression is fairly similar  to the one we obtained, namely eqs. (\ref{MS bar input}) and (\ref{massive input}) ; it produces similiar effects when z goes to zero or one.

The present approach differs from the BLL study of ref [2] in which the input $D^{\gamma,\overline{MS}}(z,M_0^2)$ is equal to zero. This leads to a different behavior of $D_g^\gamma(z,M^2)$ at small values of $z$ that we discuss in the next section.

\section{Numerical Studies of the Anomalous Component} 

In this section, we perform a numerical study of the anomalous fragmentation function. Later we shall add the VDM contribution in order to obtain the complete fragmentation functions. We solve the DGLAP equation in which the kernels are massless and take into account the effect of the mass of heavy quarks by using thresholds at $\mu^2=m_c^2$ and $\mu^2=m_b^2$ with $D_c^\gamma(\mu^2<m_c^2)=0$ and $D_b^\gamma(\mu^2<m_b^2)=0$. Then, the input (\ref{massive input}) is introduced at $m_c$ and $m_b$. We use the following values: $\Lambda_{QCD}^{(4)}=230\ MeV$, $m_{charm}=1.5\ GeV$ and $m_{bottom}=4.5\ GeV$.

\begin{xmgrfigure}
\centering
\includegraphics[angle=-90,width=\epswidth]{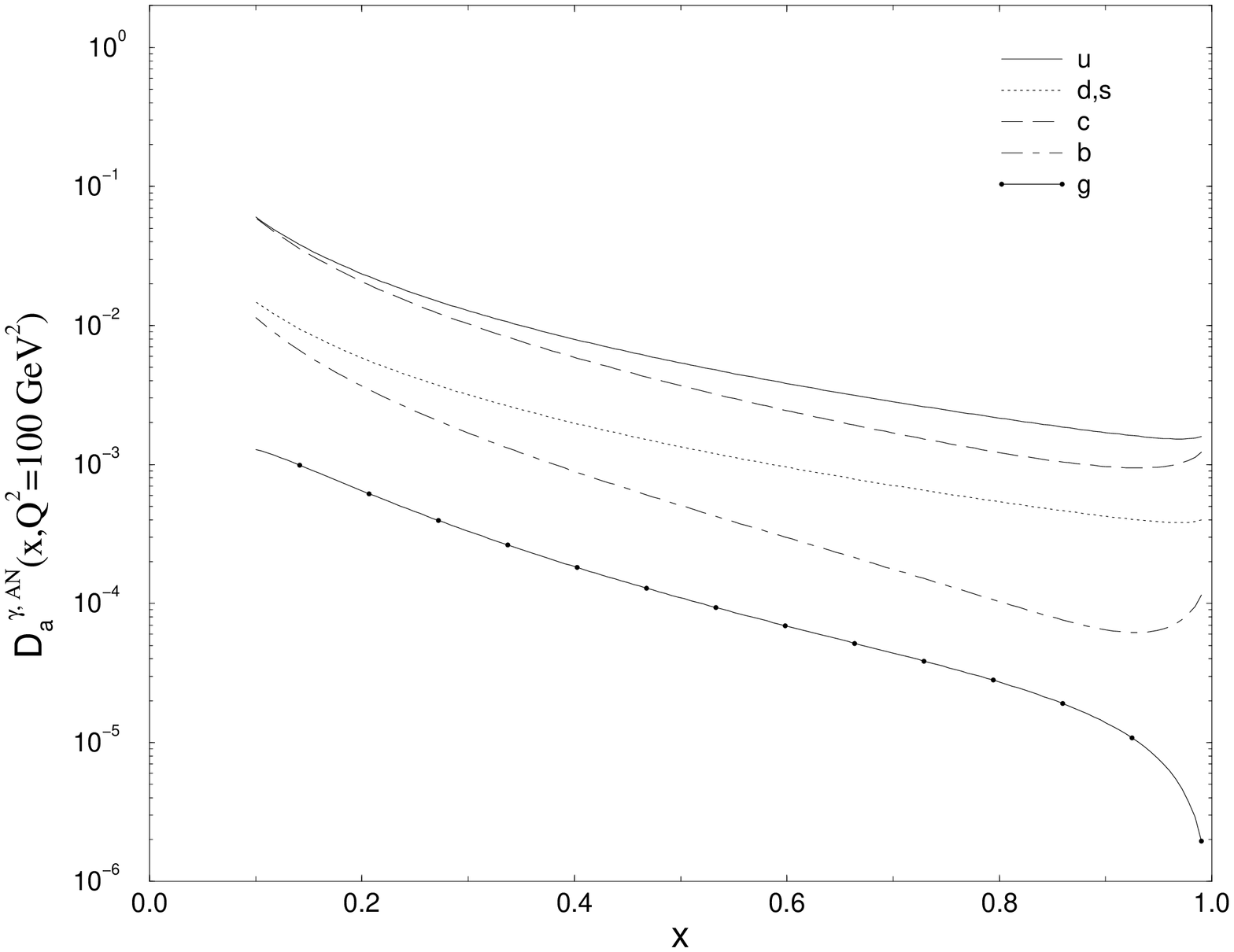}
\caption{\it{Anomalous component with $D^{\gamma , AN}(z, M_0^2= 0.5\ GeV^2)=
D^{\gamma , \overline{MS}}(z)$}}
\label{AN input MSbar}
\end{xmgrfigure}
\begin{xmgrfigure}
\centering
\includegraphics[angle=-90,width=\epswidth]{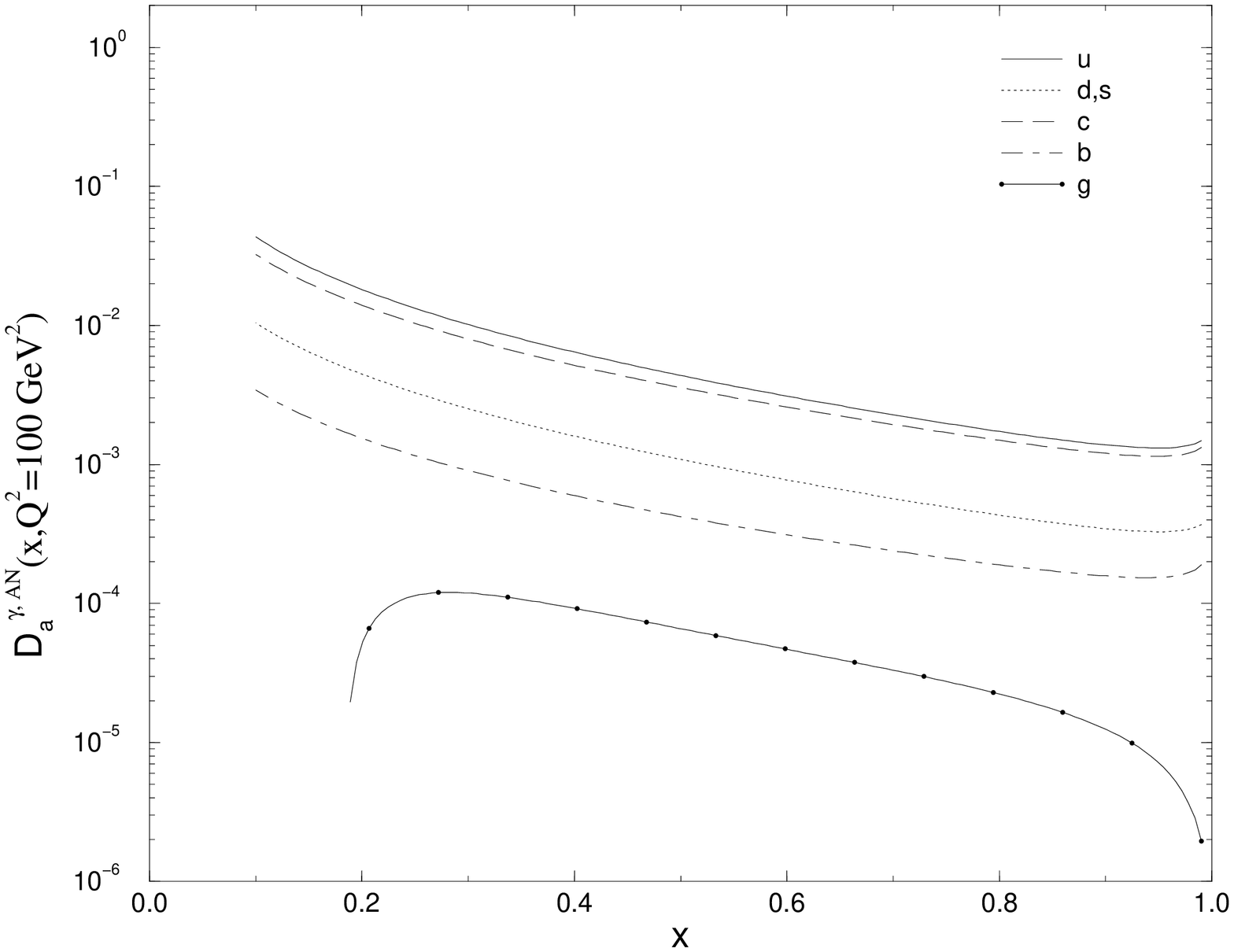}
\caption{\it{Anomalous component with $D^{\gamma , AN}(z, M_0^2= 0.5\
GeV^2)=0$}}
\label{AN input nul}
\end{xmgrfigure}

\begin{xmgrfigure}
\centering
\includegraphics[angle=-90,width=\epswidth]{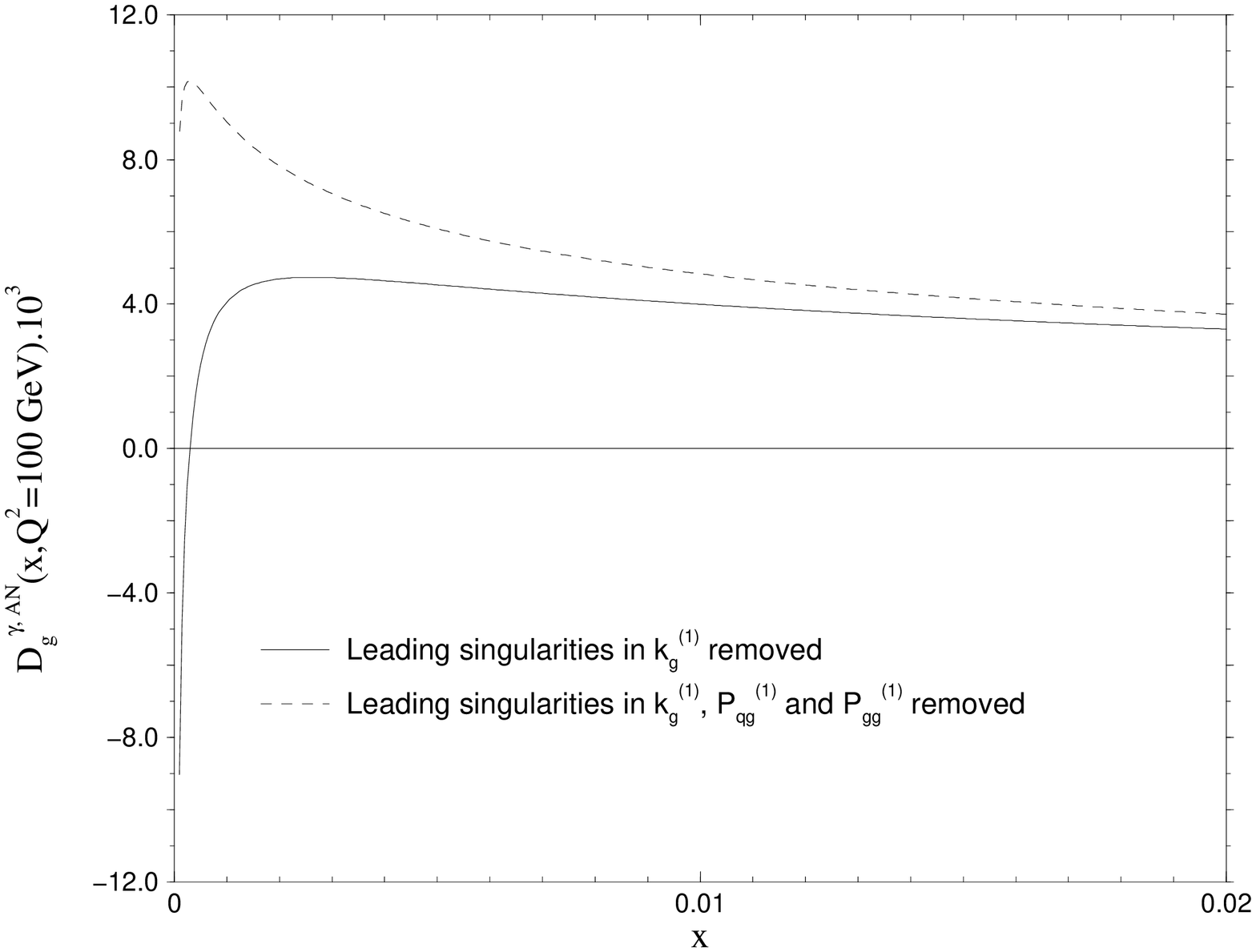}
\caption{\it{Comparison of the anomalous gluon fragmentation functions
with a null input at $Q_0^2 = 0.5\ GeV^2$ and various singularities
removed from kernels.}}
\label{Singular}
\end{xmgrfigure}

In fig.\ 1 we display the anomalous fragmentation functions obtained with $M_0^2 =0.5\ GeV^2$ and the input $D^{\gamma , \overline{MS}}(z, M_0^2)$, whereas in fig.\ 2 we show the results obtained with the boundary condition $D^{\gamma,AN}(z, M_0^2) = 0$ ($D^{\gamma , \overline{MS}} = 0$). The effects of $D^{\gamma, \overline{MS}}$ are important at small values of $z$, especially in the gluon case. In both figures, the gluon fragmentation function is negative at small $z$. But the $z$-range in which the $D^{\gamma, AN}_g$ is positive is larger with the $D^{\gamma, \overline{MS}}$ input.

This small-$z$ behavior of $D_g^{\gamma}(z, M^2)$ is due to BLL corrections to the LL solution which does not show such a pattern. The BLL kernels have a singular behavior at small~$z$~:
\bea
K_{\gamma g}^{(1)}(z) &\sim&   
{T_R \over 2} \left ( {16 \over 3} \ {\ln z \over z} \right ) \nonumber \\
P_{gq}^{(1)}(z) &\sim& \left \{ 2N_f\ C_F \ 
C_G \left ( - 4 {\ln^2 z \over z} \right ) + O \left ( {\ln z \over z} \right )
\right \} \nonumber \\
P_{gg}^{(1)}(z) &\sim& \left \{ C_G^2 \left ( - 4
{\ln^2 z \over z} \right ) + O\left ( {\ln z \over z} \right ) \right \}.
\label{kernels singularities}
\eea
\noindent The effect of the BLL inhomogeneous kernel $K_{\gamma g}^{(1)}$ is particularly important, because the Leading Order term vanishes ($K_{\gamma g}^{(0)} = 0$). If we drop the most singular term (\ref{kernels singularities}) of $K_{\gamma g}^{(1)}(z)$, we obtain a gluon fragmentation function which becomes negative only at very small values of $z$ ($z < 3.10^{-4}$) (fig.\ 3) where the effect of the homogeneous kernels is important. When the singular behavior of $P_{gg}^{(1)}$ and $P_{gq}^{(1)}$ are also removed, the fragmentation function is positive.

The $z$-domain in which the singular parts of the kernels are important has not been explored by experiment. At LEP, we have $z \gtrsim .7$ and in large-$p_{\bot}$ experiments $<z> \simeq .5$, far from the region where $D_g^{\gamma}(z, M^2)$ is negative. Therefore it is not necessary to treat this small-$z$ region more carefully by resumming to all orders the singular terms (\ref{kernels singularities}). 

When $z \to 1$, the kernels are also singular and the quark fragmentation functions are dominated by the BLL inhomogeneous contribution
\beq D_q^{\gamma}(n, M^2) \sim - {\alpha \over 2 \pi} e_q^2\ {K_{\gamma
q}^{(1)}(n) \over P^{(0)}(n)} \mathrel{\mathop {\sim}^{n \to \infty}} {\alpha
\over 2 \pi} e_q^2\ {C_F \ \ln^2 n/n \over 2 C_F \ \ln \ n} = { \alpha \over 4
\pi} e_q^2\ {\ln \ n \over n} \label{asymptotic z=1}  \eeq 
\noindent showing that
\beq
D_q^{\gamma}(z, M^2) \sim {\alpha \over 2 \pi} \ln {1 \over 1 - z}. 
\eeq
\noindent In the cross section (\ref{cross section}) this logarithmic term is cancelled by contribution coming from $C_q^{(1)}$ and $C^{\gamma}$~; as a result, the cross-section is regular when $z \to 1$.

\section{Vector Dominance Model and \\Non-Perturbative Input}

In the Vector Dominance Model, the photon is described by a superposition of vector mesons (we neglect the $J/\psi$ contribution)
\beq
\gamma = {g \over \sqrt{2}} \left ( \rho + {\omega \over 3} - {\sqrt{2} \over 3}
\phi \right ) = g \left ( {2 \over 3} (u\bar{u}) - {1 \over 3} (d \bar{d} ) - 
{1 \over 3} (s \bar{s}) \right ) 
\label{gamma-rho}
\eeq
\noindent where $g^2 \simeq \alpha$. In $e^+e^-$-annihilation, the final quark (or antiquark) first fragments into a vector meson (or a $(q\bar{q})$ state of spin 1) which is coupled to a photon through (\ref{gamma-rho}). From (\ref{gamma-rho}) we obtain
\beq
D_q^{\gamma , VDM} = g^2 \left ( {4 \over 9} D_q^{(u\bar{u})} + {1 \over
9}D_q^{(d\bar{d})} + {1 \over 9} D_q^{(s\bar{s})} \right ) 
\label{vdm}
\eeq
\noindent where $D_q^{(n\bar{n})}$ is the fragmentation function of quark $q$ into the $(n\bar{n})$ bound state. We assume that the fragmentation of the quark q into the $n\overline{n}$ bound state is given by the fragmentation into a $\rho$-meson~:
\bea
D_q^{(q \bar{q})} &=& 2 D_q^{{\rho}^0,v} + D_q^{{\rho}^0,s} \nn \\
D_{q'\neq q}^{(q \bar{q})} &=& D_{q'}^{\rho^0,s}.
\eea 
\noindent $D_q^{{\rho}^0 ,v}$ is the ``valence'' part for which the $\rho^{0}$-meson contains the quark $q$ and $D_{q}^{{\rho}^0,s}$  is the ``sea'' part for which the quark $q$ does not enter the meson. The factor 2 comes from the SU(3) wave function of the $\rho^0$-meson. We can express the VDM fragmentation function (\ref{vdm}) in terms of the quark and gluon fragmentation into $\rho^0$-meson. 

We use data from ALEPH \cite{aleph} and HRS \cite{hrs} ($\sqrt s = 29\ GeV$) in order to constrain $D_q^{\rho}(z, M^2)$ and $D_g^{\rho}(z, M^2)$. We found that data from MARK II \cite{markII}, TASSO \cite{tasso} and DELPHI \cite{delphi} are not compatible with those from ALEPH and HRS. Because HRS has the greatest statistics, we have chosen the latter. Since data from JADE \cite{jade} does not add constraints, they was not taken into account.

We use the following parametrization of the fragmentation functions at $Q_0^2=2\ GeV^2$ for the gluon and the quarks up, down, strange and charm and at $Q_0^2=m_b^2\ GeV^2$ for the quark bottom~:

\bea
&D_u^{\rho,v}(x)=&D_d^{\rho,v}(x)=N_V\ x^{\alpha_V}(1-x)^{\beta_V} \nn	\\
&D_u^{\rho,s}(x)=&D_d^{\rho,s}(x)=D_s^{\rho,s}(x)=
	N_S\ x^{\alpha_S}(1-x)^{\beta_S}	\nn			\\
&D_c^{\rho,s}(x)=&N_c\ x^{\alpha_c}(1-x)^{\beta_c} 			\\
&D_b^{\rho,s}(x)=&N_b\ x^{\alpha_b}(1-x)^{\beta_b} \nn			\\
&D_g^{\rho,s}(x)=&N_g\ x^{\alpha_g}(1-x)^{\beta_g} \nn
\eea

We reduced the number of free parameters in order to avoid too strong a correlation between them. We make the assumption that the behavior of the $c$ and $b$ quarks is related as follows~: $\alpha_b=\alpha_c$ and $\beta_b=\beta_c+2$. Furthermore, the exponents $\alpha_a$ are fixed. First, we found that it is not possible to fit HRS and ALEPH data if we keep the HRS point at $x=0.652$. For this reason, we made the fits without this point. Then, in a first fit, we fixed $\beta_g$ (set I).  When comparing the ratios $D_g^{\pi^0}/D_u^{\pi^0}$ (we used results from ref \cite{BKK}) and $D_g^{\rho}/D_u^{\rho}$, we noticed that the former is bigger by a factor of 3 to 10 (depending on the value of $x$ and $Q^2$) than the latter. Because the difference between non-perturbative mechanisms of fragmentation into $\rho^0$ or $\pi^0$ should be reduced in these ratios, they should be of the same order. Therefore our gluon fragmentation function which is not well constrained by our $e^+e^-$ data is probably too small. Thus we performed a second fit for which we fixed $N_g$ in order to obtain a ratio  $D_g^{\rho}/D_u^{\rho}$ of the order of the same ratio for pion (set II). The values of the parameters are shown in Table \ref{table sets}. We can see that the increase of $N_g$ implies a decrease of the normalization for the heavy quarks. We plot on figures \ref{fig aleph} and \ref{fig hrs} the comparison between fitted data and computed cross sections. The $\chi^2_{dof}$ is equal to 1.33 for set I and 1.22 for set II.

\begin{xmgrfigure}
\centering
\includegraphics[
angle=-90,width=\epswidth]{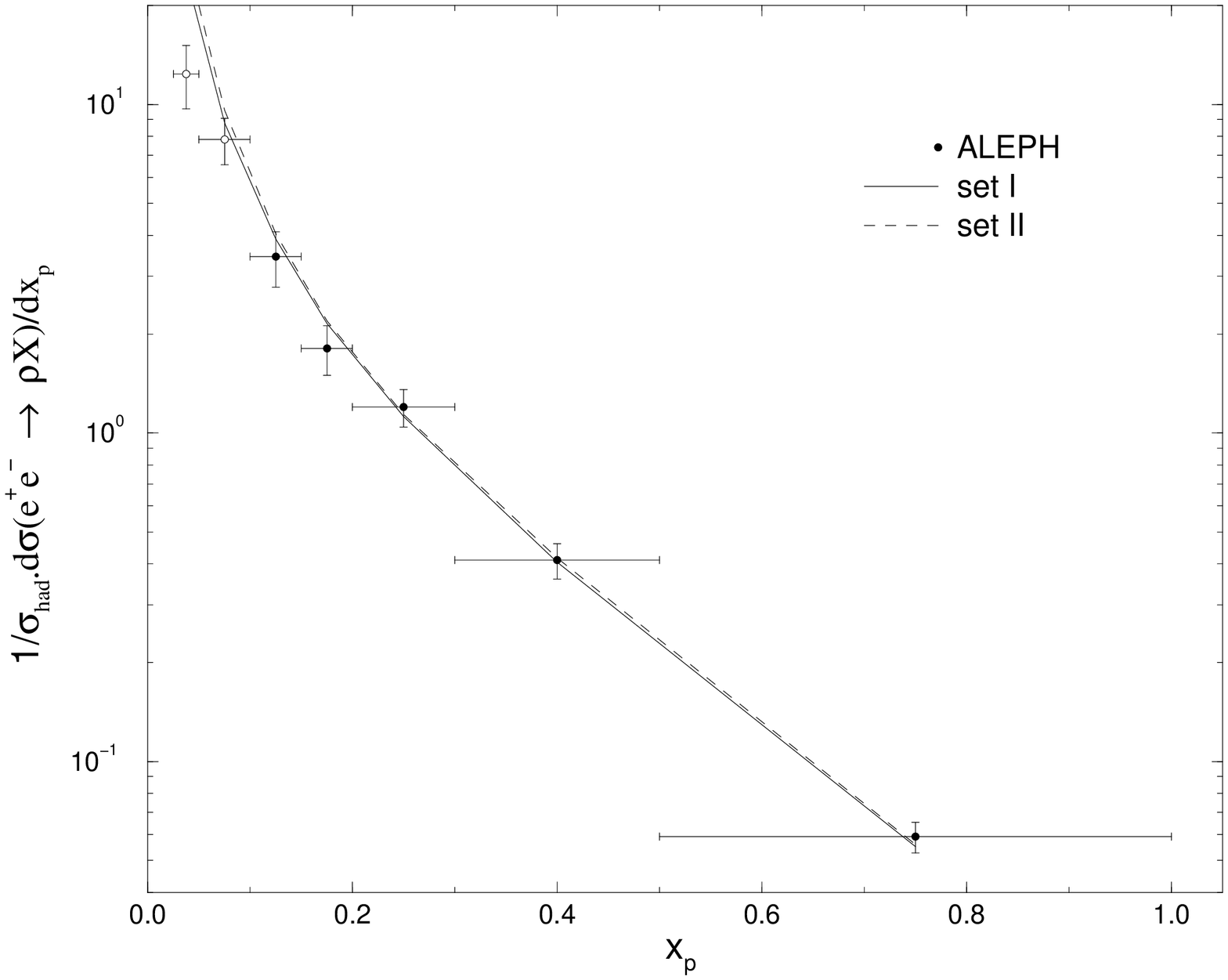}
\caption{\it{Comparison between ALEPH data and predictions corresponding 
to set I and II. Black dots correspond to points used in the fits.}} 
\label{fig aleph}
\end{xmgrfigure}

\begin{xmgrfigure}
\centering
\includegraphics[
angle=-90,width=\epswidth]{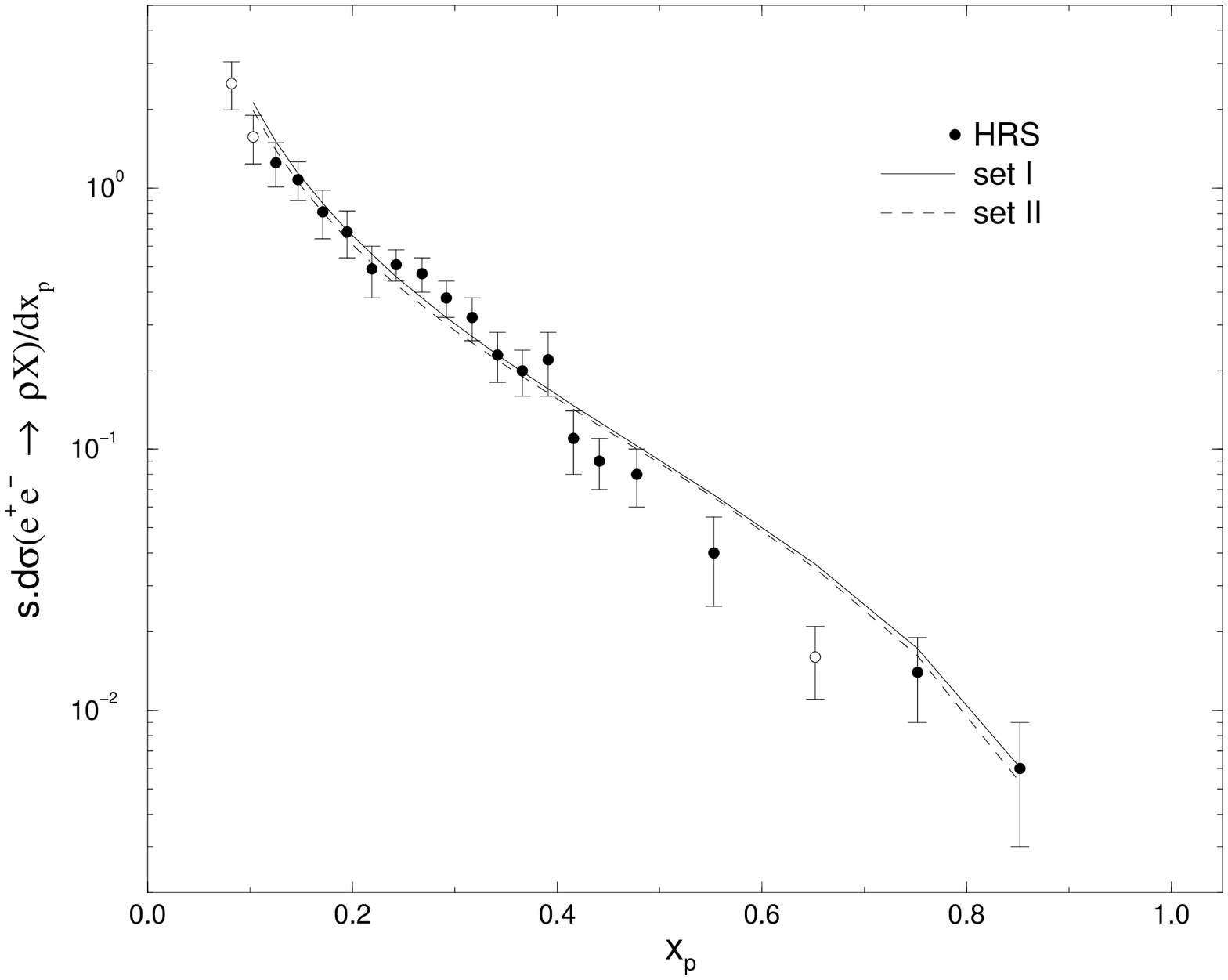}
\caption{\it{Comparison between HRS data and predictions corresponding 
to set I and II. Black dots correspond to points used in the fits.}} 
\label{fig hrs}
\end{xmgrfigure}

\begin{table}[b]
\centering
\begin {tabular}{|l|l|l|l|}
\hline
\multicolumn{4}{|c|}{set I} \\
\hline
		&N		&$\alpha$	&$\beta$	\\
\hline
valence (u,d)	&0.785 		&-0.5 		&1.499 		\\
sea (u,d,s)	&0.111 		&-1		&2.912 		\\
c		&0.567		&-1  		&5.502 		\\
b		&1.020		&-1  		&7.502 		\\
g		&0.108 		&-1  		&3 		\\
\hline
\end{tabular}
\begin {tabular}{|l|l|l|l|}
\hline
\multicolumn{4}{|c|}{set II} \\
\hline
		&N		&$\alpha$	&$\beta$	\\
\hline
valence (u,d)	&1.140  	 &-0.2	   	&1.693	 	   \\
sea (u,d,s)	&0.100	 	 &-0.3	   	&3.437	 	   \\
c		&0.132	  	 &-1	   	&4.820	 	   \\
b		&0.103	  	 &-1	   	&6.820	 	   \\
g		&2.550	 	 &-0.3	   	&3	   	   \\
\hline
\end{tabular}
\caption{\it{Fitted parameters for the fragmentation functions to $\rho$. The exponents $\alpha_a$ are fixed.}}
\label{table sets}
\end{table}

\section{Full Fragmentation Functions and Comparison with Experiment}

We obtain the complete fragmentation functions by adding the VDM contributions to the anomalous contributions. They are given in fig.\ \ref{frag100_I}, \ref{frag100_II} (for $M^2 = 100\ GeV^2$) and fig.\ \ref{frag10000_I}, \ref{frag10000_II} (for $M^2 = 10^4\ GeV^2$).

\suppressfloats

The results of fig.\ \ref{frag100_I} and \ref{frag10000_I} correspond to set I of parton into $\rho$-meson fragmentation functions discussed in the preceeding section, whereas those of fig.\ \ref{frag100_II} and \ref{frag10000_II} correspond to set II. We notice a sizeable difference only for the gluon fragmentation functions; in this case the VDM contributions are very different. On the other hand the VDM contributions to the quark fragmentation functions are small and the curves of fig.\ \ref{frag100_I} to  \ref{frag10000_II} are very similar to the corresponding curves of fig.\ \ref{AN input MSbar}. These distributions can be compared with those we obtained in ref [2] (fig.\ 10 to 13). The latter are larger at small $z$ and not too large $Q^2$, the difference being essentially due to a different VDM input. In ref [2], we assume that the fragmentation in $\rho$-meson is similar to the fragmentation in $\pi^0$ and we use the distributions $D_a^{\pi^0}$ of ref [7] as VDM input. In this case, the input we obtained after a fit to data is much smaller.

\begin{xmgrfigure}
\centering
\includegraphics[angle=-90,width=\epswidth]{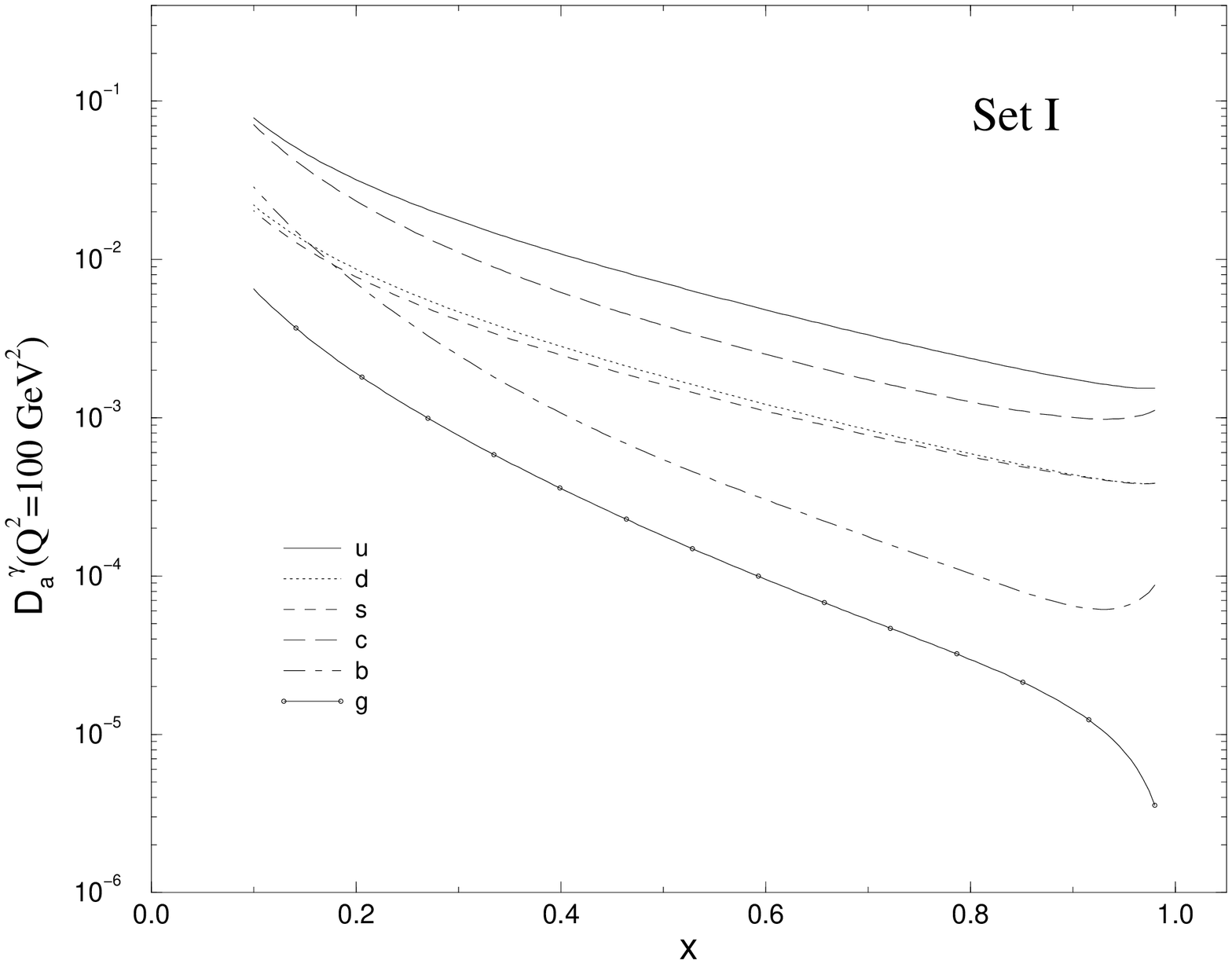}
\caption{\it{The fragmentation functions at $Q^2 = 100\ GeV^2$.}} 
\label{frag100_I}
\end{xmgrfigure}

\begin{xmgrfigure}
\centering
\includegraphics[angle=-90,width=\epswidth]{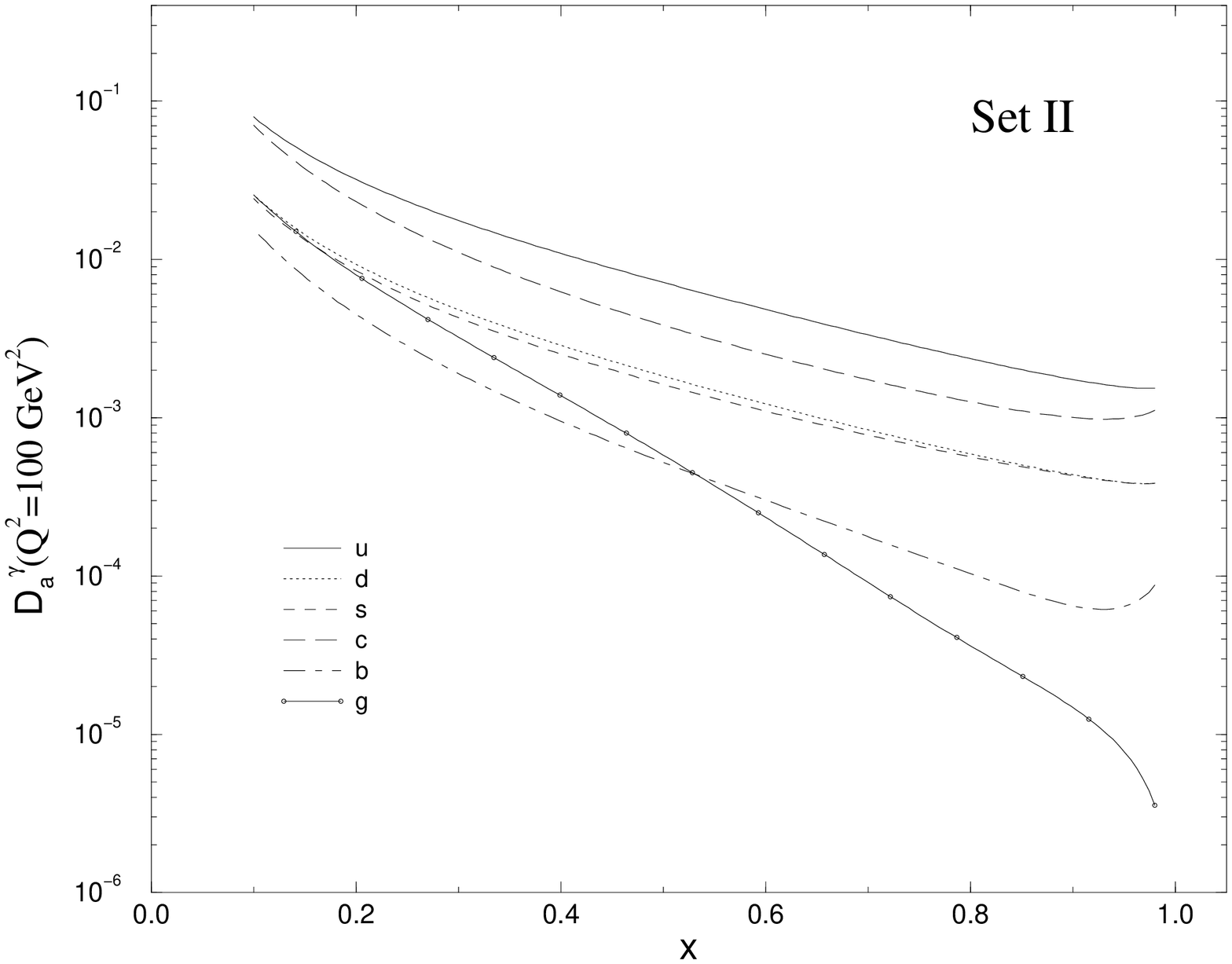}
\caption{\it{The fragmentation functions at $Q^2 = 100\ GeV^2$.}} 
\label{frag100_II}
\end{xmgrfigure}

\begin{xmgrfigure}
\centering
\includegraphics[angle=-90,width=\epswidth]{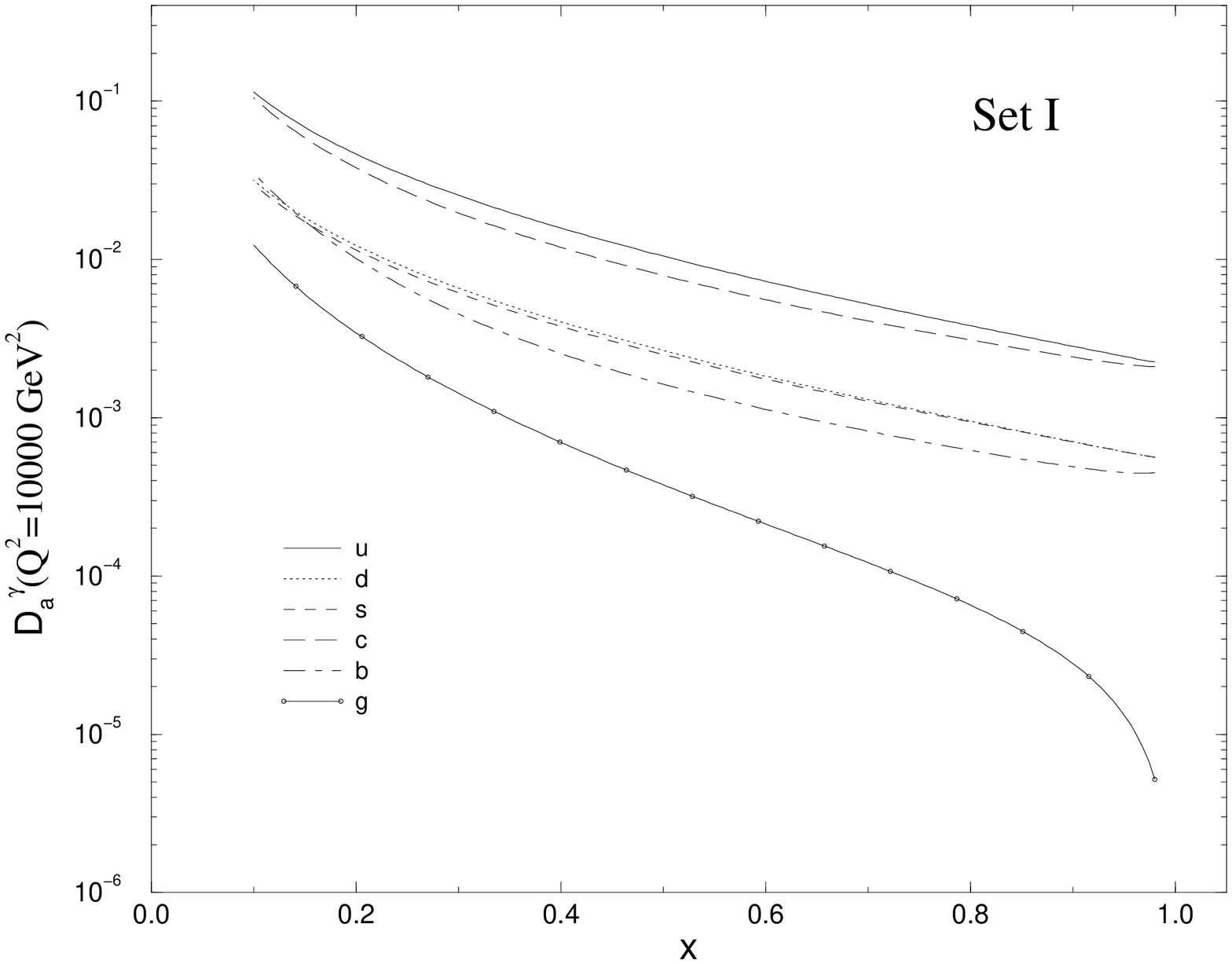}
\caption{\it{The fragmentation functions at $Q^2 = 10000\ GeV^2$.}} 
\label{frag10000_I}
\end{xmgrfigure}

\begin{xmgrfigure}
\centering
\includegraphics[angle=-90,width=\epswidth]{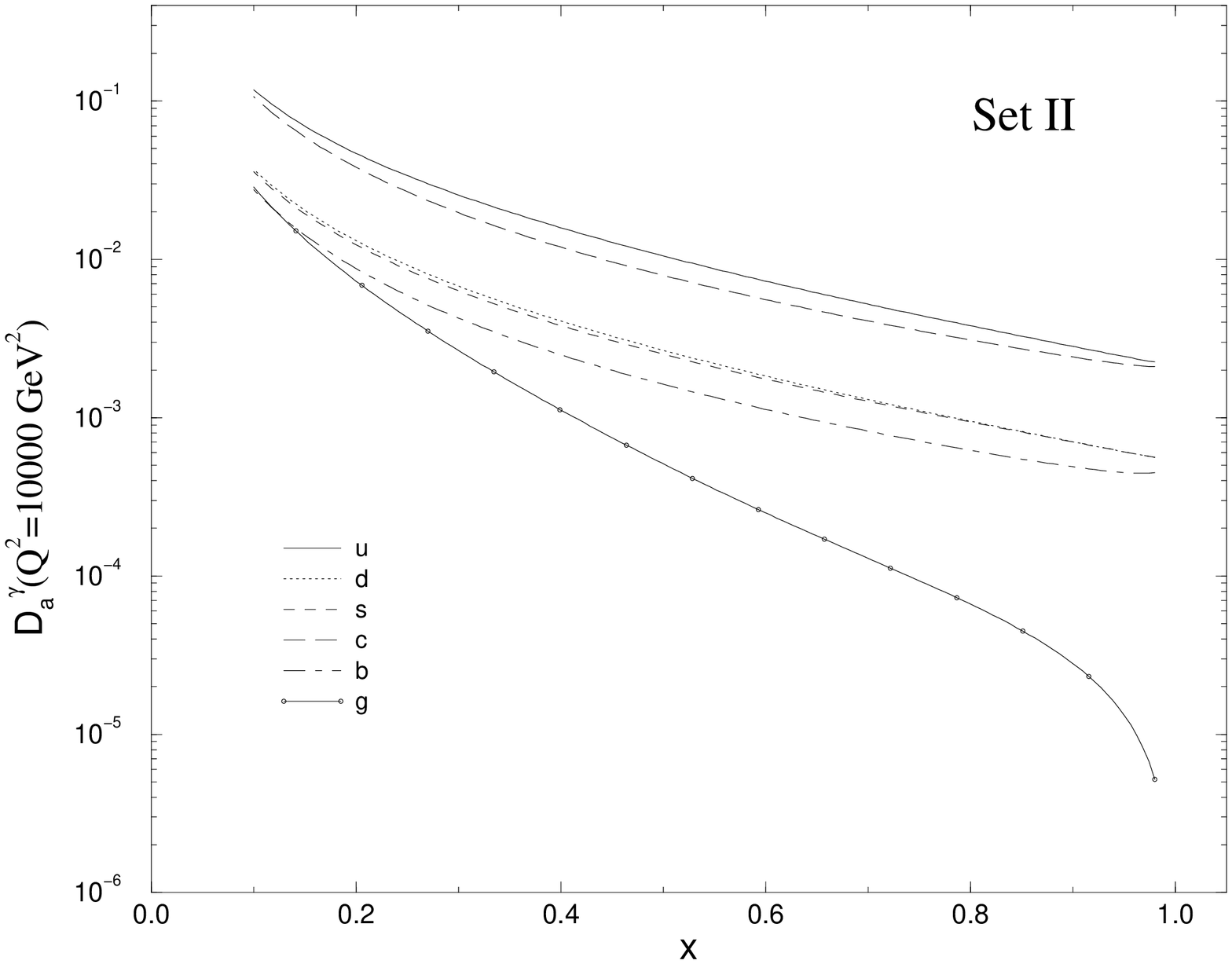}
\caption{\it{The fragmentation functions at $Q^2 = 10000\ GeV^2$.}} 
\label{frag10000_II}
\end{xmgrfigure}

In fig.\ \ref{Duke-Owens}, we compare our results with the LL parametrization of Duke and Owens [12]. However one must keep in mind that BLL distribution functions are factorization scheme dependent, and that our distributions are calculated in the $\overline{MS}$ scheme. A better comparison is provided by the cross section $d\sigma^{\gamma}/dz$, an invariant observable which can be compared to experiment.

\begin{xmgrfigure}
\centering
\includegraphics[angle=-90,width=\epswidth]{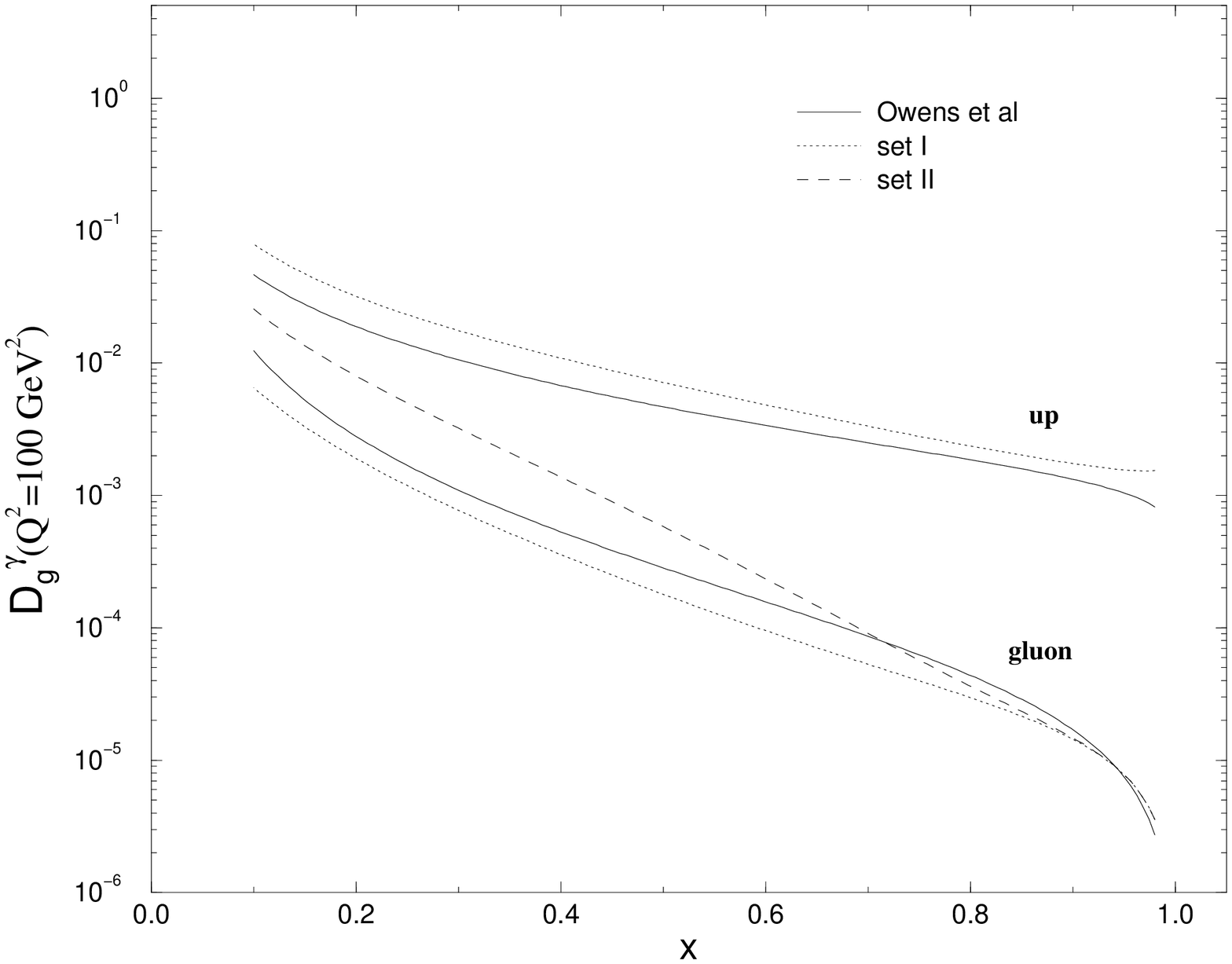}
\caption{\it{The fragmentation function for gluon and quark up of our sets of fragmentation functions compared with those of Duke-Owens at $Q^2=100\ GeV^2$. For the quark, set I and set II cannot be distinguished.}} 
\label{Duke-Owens}
\end{xmgrfigure}

At present, there is no data with which to compare. ALEPH data could seem to be a basis for comparison, but it is produced by an analysis in jets. Following  [3], this Collaboration defines the fragmentation into a photon within a jet. The fragmentation function $D_{jet}^{\gamma}$ defined in that way does not correspond to the functions calculated in this paper which are fully inclusive~; we do not put any limitation on the phase-space of the hadrons which accompany the photon.

In order to better understand the difference between ALEPH results and our predictions, let us consider the decay of a Z-boson (of momentum $Q$) into a photon ($p_1$), a quark ($p_2$) and an anti-quark ($p_3$). We define $z_i=2p_i.Q/Q^2$ and $y_{ij}=2p_i.p_j/Q^2$, where $z_1$ is the inclusive photon fragmentation variable. We have $1-z_i=y_{jk}$ (i,j,k different) and $\sum_{i<j} y_{ij}=1$. The variable used by ALEPH to describe the photon in the jet (here the jet is made of the photon and the quark) is $z_\gamma=z_1/(z_1+z_2)=z_1/(1+y_{12})$. However an integration is performed on $y_{12}$ within the jet, so that the effective value $z_1^{eff}$ at which one should compare our results is larger than $z_\gamma$. But if we assume that the largest contribution to the integral comes from the collinear region $y_{12} \approx 0$, we obtain $z_1 \approx z_\gamma$. Hereafter we use this assumption.

One must also notice that the ALEPH Collaboration uses the Durham algorithm [28] to define a jet. According to this algorithm, $y_{12}^{max} = (1-z_\gamma)/(1+z_\gamma)$; therefore the scale in the fragmentation function is no longer $Q^2$, but $(1-z_\gamma)/(1+z_\gamma) Q^2/z_\gamma$ (the extra $1/z_\gamma$ comes from the fact that in the inclusive case $y_{12}^{max}=z_1 Q^2$ and this factor $z_1 \approx z_\gamma$ is already included in our calculation). 

Finally, one must keep in mind that higher order QCD corrections to the $q\overline{q}\gamma$ process can generate logarithms of the jets parameter (e.g. $y_{cut}$) coming from the limited phase space integration, which are not present in the fully inclusive case. However for  $z_\gamma$ large enough, $y_{cut}$ does no longer constrain the phase space and the comparison between our predictions and ALEPH results is not spoiled by this effect.

This comparison is shown in fig.\ \ref{eeToGamma} for 2-jets events and $y_{cut}=0.1$. We display our predictions for two scales in order to exhibit their sensitivity to the latter. The agreement is quite satisfactory. It is interesting to notice that, in this $z_\gamma$-region, one essentially tests the anomalous component of the fragmentation functions; once $Q_0^2$ is chosen, these parts are a pure prediction of the perturbative QCD. $Q_0^2$, which is of the order of $m_\rho^2$, caracterizes the border between the perturbative and non perturbative regions (cf eq.(6)).

\begin{xmgrfigure}
\centering
\includegraphics[angle=-90,width=\epswidth]{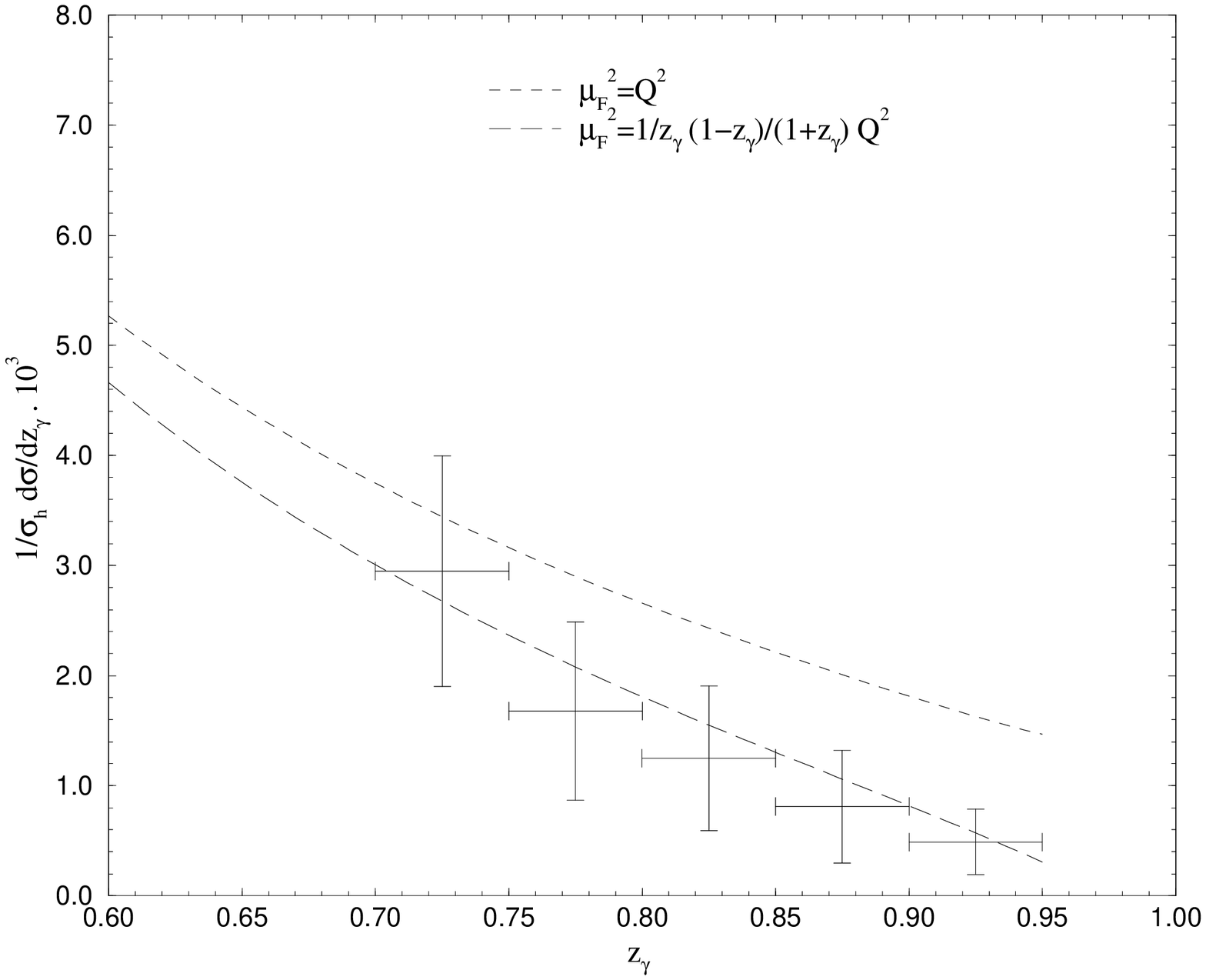}
\caption{\it{Comparison between ALEPH data and our predictions for the direct production of photon.}} 
\label{eeToGamma}
\end{xmgrfigure}

\section{Conclusions}

We have studied the parton to photon fragmentation functions beyond the leading order. We recalled that the traditional decomposition of this functions in non-perturbative and anomalous parts depends on the photonic factorization scheme. Performing a careful analysis of this dependence, we propose a new definition of the perturbative and non perturbative components. In this approach, all the scheme dependence is put in the perturbative part. By using a VDM approach, we constrain the non perturbative component of the fragmentation function that we deduced from parton to rho fragmentation functions. The latter was obtained from a fit to LEP and PEP data. Finally, we propose two new sets of parton to photon fragmentation functions \footnote{FORTRAN subroutines which compute these fragmentation functions are available on request by e-mail to fontanna@qcd.th.u-psud.fr}. We used them to give new predictions for the production of direct $\gamma$ at LEP that agree well with experimental data. However, these data obtained by an analysis in jets are not fully inclusive and they do not exactly coincide with our inclusive fragmentation functions. Therefore, fully inclusive data for direct photon production would be very interesting, as they would allow to test a beautiful prediction of perturbative QCD.


\clearpage

\begin{thebibliography}{99}

\bibitem{1} E.Witten, Nucl. Phys. {\bf 210} (1977) 189.
Actually Witten obtained his results for the crossed reaction, the Deep Inelastic $\gamma \gamma^*$ Scattering, by using the operator-product-expansion method.  
\bibitem{2} P. Aurenche, P. Chiappetta, M.Fontannaz, J. Ph. Guillet and E. Pilon,  Nucl. Phys. {\bf B399} (1993) 34.   
\bibitem{3} D. Buskulic et al., ALEPH Collaboration, Z. Phys. {\bf C69} (1996) 
365.   
\bibitem{4} E706 Collaboration, T. Ferbel in ``Proceedings of the Rencontres de 
Moriond'', March 1996, edited by J.Tran Thanh Van. 
\bibitem{5} UA6 Collaboration, private communication from M.Werlen.  
\bibitem{delphi} P. Abreu et al., DELPHI Collaboration, Z. Phys. {\bf C65}
(1995) 587.
\bibitem{aleph} D. Busculic et al., ALEPH Collaboration, Z. Phys. {\bf C69}
(1996) 379.
\bibitem{8} P. Aurenche, M.Fontannaz, J.-Ph. Guillet, Z.Phys. {\bf C64} (1994)
 621. 
\bibitem{9} G. Kramer and B. Lampe, Phys. Lett. {\bf B269} (1991) 401\\
E.W.H. Glover and W.J. Stirling, Phys. Lett. {\bf B295} (1992) 128\\ 
Z.Kunst and Z. Troczanyi, Nucl. Phys. {\bf B394} (1993) 139
\bibitem{10} E.L. Berger, X. Guo and J. Qiu, Phys. Rev. Lett. {\bf 76} (1996) 
2234  
\bibitem{11} P. Aurenche, M. Fontannaz, J.P. Guillet, A. Kotikov, E. Pilon,
Phys. Rev. {\bf D55} (1997) R 1124
\bibitem{12} C. H. Llewellyn Smith, Phys. Lett. {\bf 79B} (1977) 189.\\
K. Koller, T. F. Walsh and P. M. Zerwas, Z. Phys. {\bf C2} (1979) 197.\\
A. Nicolaidis, Nucl. Phys. {\bf B163} (1980) 156.\\
D. W. Duke and J. F. Owens, Phys. Rev. {\bf D26} (1982) 1600.
\bibitem{13} We follow the notations of [2]. 
\bibitem{14} G. Curci, W. Furmanski and R. Petronzio, Phys. Lett. {\bf B175} 
(1980) 27.\\
W. Furmanski and R. Petronzio, Phys. Lett. {\bf B97} (1980) 437. 
\bibitem{15} W. Furmanski and R. Petronzio, Z. Phys. C (198 ). 
\bibitem{16} In this paper, we use a FORTRAN program written by P. Nason and 
modified to take into account the inhomogeneous kernels. 
\bibitem{17} M. Gl\"uck, E. Reya and A. Vogt, Phys. Rev. {\bf D48} (1993) 116.

The moments of the kernels can be found in this reference.
\bibitem{18} W. A. Bardeen and A. J. Buras, Phys. Rev. {\bf D20} (1979) 166 
derived expression (8) for the DIS case.
\bibitem{19} G. Altarelli, R. K. Ellis, G. Martinelli,
S.Y. Pi, Nucl. Phys. {\bf 160} (1979) 301. 
\bibitem{20} An all order discussion of the FS dependence can be found in 
appendix B of [8]. 
\bibitem{21} P. Nason, B. R. Webber, Nucl. Phys. {\bf B421} (1994) 473.
\bibitem{hrs} S. Abachi et al., HRS Collaboration, Phys. Rev. {\bf D40} (1989)
706.
\bibitem{markII} H. Schellman, MARKII Collaboration, SLAC-PUB-3448 (1984).   
\bibitem{tasso} W. Brandelik et al., TASSO Collaboration, Phys. Lett. {\bf 117B}
(1982) 135.
\bibitem{jade} W. Bartel et al., JADE Collaboration, Phys. Lett. {\bf 145B}
(1984) 441.
\bibitem{BKK} J. Binnewies, B.A. Kniehl, G. Kramer, Phys. Rev. {\bf D52}(1995) 
 4947 
\bibitem{27} P. Chiappetta, M. Greco, J.P. Guillet, S. Rolli, M. Werlen, Nucl.
Phys. {\bf B412} (1994) 3
\bibitem{28} Y.Dokshitzer, Z.Phys. {\bf C17} (1991) 1441
\end{thebibliography}
\end{document}